\documentclass[preprint,preprintnumbers,amsmath,amssymb,nofootinbib,aps]{revtex4}
\usepackage[utf8]{inputenc}
\usepackage{amsmath}
\usepackage{amsfonts}
\usepackage{amssymb}
\usepackage{color}
\pdfoutput=1
\usepackage{epsfig}
\usepackage{graphicx}
\usepackage{bm}

\begin{document}

\title{Fermionic wave functions and Grassmann fields as possible sources of dark energy}

\author{L. C. T. Brito$^{1}$}\email{lcbrito@ufla.br}
\author{S. H. Pereira$^{2}$} \email{s.pereira@unesp.br}
\author{L. N. Barboza$^{1,5}$}\email{lauziene.barboza@gmail.com}
\author{J. C. C. Felipe$^{3}$}\email{jean.cfelipe@ufvjm.edu.br}
\author{J. F. Jesus$^{4,2}$}\email{jf.jesus@unesp.br}

\affiliation{$^1$Departamento de F\'isica, Intituto de Ci\^encias Naturais, Universidade Federal de Lavras (UFLA), Caixa Postal 3037, 37200-900, Lavras, MG, Brasil,\\
$^2$Departamento de F\'isica, Faculdade de Engenharia e Ci\^encias de Guaratinguet\'a, Universidade Estadual Paulista (UNESP),   Av. Dr. Ariberto Pereira da Cunha 333, 12516-410, Guaratinguet\'a, SP, Brazil,\\
$^3$Instituto de Engenharia, Ci\^encia e Tecnologia, Universidade Federal dos Vales do Jequitinhonha e Mucuri (UFVJM), Avenida Um, 4050 - 39447-790 - Cidade Universit\'aria - Jana\'uba - MG- Brazil,\\
$^4$Instituto de Ciências e Engenharia, Universidade Estadual Paulista (UNESP) - R. Geraldo Alckmin, 519, 18409-010, Itapeva, SP, Brazil,\\
$^5$Instituto de F\'isica, Universidade Federal Fluminense, Avenida General Milton Tavares de Souza s/n, Gragoat\'a, 24210-346 Niter\'oi, RJ, Brazil.
}

\begin{abstract}
We study a cosmological model with a fermionic field which can be interpreted as a source of dark energy in the universe. Two different approaches were considered, the first one with a massless fermionic field represented by a standard wave-function and the second one where a massive field is a Grassmann variable. {The first case naturally reduces to a XCDM model with a constant equation of state parameter, while the last case reproduces a $w(z)$CDM model for a massive field}, and in the massless limit, the intrinsic grassmannian property of the field leads always to a vacuum equation of state parameter, irrespective the specific form of the potential. Both cases leads to a dark energy contribution of the fermionic sector. The models are totally compatible with recent cosmological data from Supernovae, BAO and Hubble parameter measurements. A brief study of linear evolution of density perturbations shows that some of the small scale problems related to standard model can be at least alleviated. 
\end{abstract}

\maketitle

\section{Introduction}

The present status of the astronomical observations of the  Cosmic Microwave Background radiation,  the large-scale distribution of galaxies, and distant Supernovae of Type Ia (SNe Ia), among others, lead us to an accurate cosmic concordance model for cosmology \cite{peebles}. The model incorporates, beyond the standard baryonic matter, a sector described by the cosmological constant $\Lambda$ (or dark energy) and the Cold Dark Matter (CDM), and is called the $\Lambda$CDM model. 

The dark sector is a fundamental part of the model, since they represent about 95\% of the total material content of the Universe, what allows the description of astronomical data accurately. On the other hand, it also brought to cosmology some of the most challenging problems of modern science. In particular, the knowledge of the nature of dark energy and dark matter is far from being settled \cite{oks,Arun}.

In the present work we review a model where the dark energy component in the cosmic evolution is a single fermionic field, permitting an alternative description of the dark energy sector of the $\Lambda$CDM. Our purpose is to discuss theoretical aspects that, as we will see,  may have cosmological consequences. The work, we believe, adds new knowledge to the numerical analysis presented in other papers \cite{Ribas:2005vr,Ribas2007}. Besides, we will highlight to what extent the model agrees with recent astronomical data \cite{Magana2018,Moresco2022,pantheon,Planck2018}.

For a starting motivation for introducing fermions in cosmology, we may look for the history of analogies in particle physics. The first unified description of states of nucleons and mesons as composed states of a fundamental fermion, for example,  was introduced in the seminal papers by Nanbu and Jona-Lasinio in analogy with superconductivity \cite{nambu}. Beyond analogy, we know now that we must interpret the Nambu-Jona-Lasinio model as an effective quantum field theory, valid at sufficiently low energy \cite{klevansky}. At present, fermions are well established as a fundamental part of the description of nature, representing quarks and leptons in the present Standard Model (SM) of elementary particles \cite{donoghue}. 

However, although we know now that the SM is the theory underlying an effective model like that proposed by Nambu and Jona-Lasinio, we do not know what fundamental theory is beyond the $\Lambda$CDM model \cite{Joyce:2014kja}. Thus, a description based on an analogy with the fields of the SM opens the possibility to explore not only the fermion component but also the scalar \cite{Steinwachs} and vector \cite{Maleknejad} degrees of freedom. A possible vector component does not concern us here because we consider an isotropic universe. But it is interesting to make some comments about models based on scalar field components and how they may be related to  fermions. Again, we look for insights into particle physics and quantum field theory.

It is well established that the dynamic of the SM is in a particular representation of the vacuum where the electroweak symmetry $SU(2)\times U(1)$ is spontaneously broken to $U(1)$. The process which explains the breaking of the gauge symmetry of the model is a second-order phase transition occurring in a quantum field theory at finite temperature. This picture is introduced in the model by the Higgs sector, described by a scalar field that in the broken phase which we observe at present gives mass to the leptons and the right massive modes correspondent to the weak vector bosons and the massless photon of the electromagnetism. Of course, it is important for cosmology since this mechanism furnish a natural description of the early universe \cite{kolbe}.

Although the scalar sector of the SM introduces problems to the dynamics that are undesirable from the physical point of view \cite{Susskind}. Among the ideas proposed to solve or at least give a better understanding of these problems,  we find the interpretation of the Higgs field as a composite of fermion fields that extend the degrees of freedom of the dynamics beyond the Standard Model \cite{Ferretti:2013kya,Barnard:2013zea,Csaki:2015hcd}. 

Recent works have studied the cosmological implications of fermionic fields in different contexts, as inflation, dark matter and dark energy \cite{Ribas:2005vr,Ribas2007,Rakhi,Vignolo,Inagaki,Channuie,Carloni}. The purpose of the present paper is to explore a possible analogy between particle physics and cosmology based on the idea that we can interpret the dark energy sector in terms of a fundamental fermion. In particular, we analyze the cosmological consequences of treating the fermion as a single wave-function or a Grassmann classical field.

The paper is organized as follows. The basic equations of the model are presented in section II. In section III, we discuss the cases in which the fermion is treated as a wave function or Grassmann field.  We consider the constraint of the models with observational data in section IV. A brief study of linear evolution of density perturbations is presented in Section V. In section VI, we present the conclusions.

\section{Fermions coupled to gravity}

We start with  the Lagrangian density of the model of Dirac fermions\footnote{In natural units.} $\psi$:
\begin{equation}
\mathcal{L}_{f}=\frac{i}{2}\left[\bar{\psi}\Gamma^{\mu}D_{\mu}\psi-\left(D_{\mu}\bar{\psi}\right)\Gamma^{\mu}\psi\right]-m\bar{\psi}\psi-V(\bar{\psi},\psi).
\label{lagrangian}
\end{equation}
where $\bar{\psi}=\psi^{\dagger}\, \gamma^{0}$, $\Gamma^{\mu}=e^{\mu}_{a}\gamma^{a}$, $m$ is a constant with dimension of mass, and $V(\bar{\psi},\psi)$ is the potential with self-interactions constructed from the bilinears $\bar{\psi}\psi$ and $i\bar{\psi}\gamma_{5}\psi$. The covariant derivative and the spin connection are \cite{Ribas:2005vr} 
\begin{eqnarray}
D_{\mu}\psi &=& \partial_{\mu}\psi-\Omega_{\mu}\psi, \\
D_{\mu}\bar{\psi} &=& \partial_{\mu}\bar{\psi}+\bar{\psi}\Omega_{\mu}, \\
\Omega_{\mu} &=& -\frac{1}{4}g_{\rho\sigma}\left[\Gamma_{\mu\nu}^{\rho}-e_{a}^{\rho}\partial_{\mu}e_{\nu}^{a}\right]\Gamma^{\sigma}\Gamma^{\nu}.
\end{eqnarray}
The $\gamma^{a}$ are the Dirac matrices, $e_{a}^{\mu}$ are the set of  tetrad defined at every point of the gravitational field, and we also have $\gamma_{5}=i\gamma^{0}\gamma^{1}\gamma^{2}\gamma^{3}$. 

The equations of motion  of the fermion  in the presence of the gravitational field are
\begin{eqnarray}
i\Gamma^{\mu}D_{\mu}\psi-m\psi-\frac{dV}{d\bar{\psi}}&=&0\nonumber\\
iD_{\mu}\bar{\psi}\Gamma^{\mu}+m\bar{\psi}+\frac{dV}{d\psi}&=&0.
\label{motion}
\end{eqnarray}
The Einstein Equations of the metric $g_{\mu\nu}$ are 
\begin{equation}
R_{\mu\nu} - \frac{1}{2} g_{\mu\nu}  R = - 8 \pi G\, T_{\mu\nu},\label{EqEinstein}
\end{equation}
where $T_{\mu\nu} = T_{m\mu\nu} + T_{f\mu\nu}$ is the total energy-momentum tensor, with $T_{m\mu\nu}$ representing the energy-momentum tensor for matter and $T_{f\mu\nu}$ for the fermion field $\psi$, obtained by:  
\begin{equation}
\delta S_{f}=-\frac{1}{2}\int d^{4}x\sqrt{-g}\,\,T_{f}^{\mu\nu}\,\delta g_{\mu\nu}.
\end{equation}

Using the Lagrangian density (\ref{lagrangian}), the variation of the action $S_{f}$ with relation to the metric $g_{\mu\nu}$    gives: 
\begin{equation}
T_{f}^{\mu\nu}=\frac{i}{4}\left[\bar{\psi}\Gamma^{\nu}D^{\mu}\psi+\bar{\psi}\Gamma^{\mu}D^{\nu}\psi-D^{\mu}\bar{\psi}\Gamma^{\nu}\psi-D^{\nu}\bar{\psi}\Gamma^{\mu}\psi\right]-g^{\mu\nu}\mathcal{L}_{f}.
\label{energia_momento}
\end{equation}
From this, we have the energy density $\rho_f$  and pressure $p_f$ of the fermion considered as a perfect fluid:
\begin{equation}
T_{f\,\,\nu}^{\mu}=g_{\nu\sigma}T_{f}^{\mu\sigma} = \left(\rho_f,-p_f,-p_f,-p_f\right).
\end{equation}
The expressions for pressure and energy density comes from the expression (\ref{energia_momento}). As a consequence of the homogeneity of the space,  we have that $\psi$ is a function only of time. It furnishes  simplified expressions for the covariant derivative:
\begin{equation}
D_{i}\psi=-\Omega_{i}\psi\,\,\,\,\,\,\,\,\,\,D_{i}\bar{\psi}=\bar{\psi}\Omega_{i}
\end{equation}
and
\begin{equation}
D_{0}\psi=\dot{\psi}\,\,\,\,\,\,\,\,\,\,D_{0}\bar{\psi}=\dot{\bar{\psi}}.
\end{equation}
Here, the ``dot" on the $\psi$  and $\bar{\psi}$ represents derivative with relation to time, and we have used the fact that $\Omega_{0}=0$ . Since 
 \begin{equation}
\Gamma^{0}=\gamma^{0},\,\,\,\,\,\,\,\,\,\,\,\,\,\,\,\,\,\,\,\,\Gamma^{i}=\frac{1}{a(t)}\gamma^{i}\,\,\,\,\,\,\,\,\,\,\,\,\text{and}\,\,\,\,\,\,\,\,\,\,\,\,\Omega_{i}=\frac{1}{2}\dot{a}(t)\gamma^{i}\gamma^{0},  
\end{equation}
we have 
\begin{equation}
\Gamma^{i}\Omega_{j}+\Omega_{j}\Gamma^{i} =0.
\end{equation}
Using this identity and the equations of motion \eqref{motion} in the expression \eqref{energia_momento} for the energy-momentum tensor we get  
\begin{equation}
p_f = \frac{1}{2}\bar{\psi}\frac{dV}{d\bar{\psi}}+\frac{1}{2}\frac{dV}{d\psi}\psi-V(\bar{\psi},\psi) \label{pressure} 
\end{equation}
and
\begin{equation}
\rho_f =  m\bar{\psi}\psi+V(\bar{\psi},\psi) \label{density}.
\end{equation}

Taking the covariant differentiation of Einstein field equations (\ref{EqEinstein}) we obtain the conservation of the total energy-momentum tensor:
\begin{equation}
    \dot{\rho}+3H(\rho + p + \bar{\omega})=0\,\label{Eqconserv}
\end{equation}
where $H=\dot{a}/a$ and $\bar{\omega}$ represents the nonequilibrium pressure when dissipative processes are taken
into account \cite{Kremer2003}. In which follows we will assume $\bar{\omega} \approx 0$.

\section{Wave function versus  Grassmann variables}
The terms with  $\frac{dV}{d\bar{\psi}}$ and $\frac{dV}{d\bar{\psi}}$ which appear in the expression \eqref{pressure} may have interesting implications  for the model. It is because the result for the derivatives depends  if we treat $\psi$  as a wave function or a relativistic classical field. In the last case,  the fermion statistic demands  the field  to be a Grassmann variable. We can consider both cases in the calculations observing that  
\begin{equation}
\frac{d\left(\bar{\psi}\psi\right)}{d\psi}=s\bar{\psi}\,\,\,\,\,\,\,\,\,\,\text{and}\,\,\,\,\,\,\,\,\,\, \frac{d\left(\bar{\psi}\gamma^{5}\psi\right)}{d\psi}=s\bar{\psi}\gamma^{5}\,,
\label{derivatives}
\end{equation}
where $s=1$ corresponds to the $\psi$ treated as a wave function, and $s=-1$ to the case that it is a classical field described by a Grassmann variable.

As pointed in \cite{Ribas:2005vr}, the most general form for the self-interaction potential $V(\bar{\psi},\psi)$ constructed from the Lorentz invariant bilinears 
$\bar{\psi}\psi$ and $i\bar{\psi}\gamma_{5}\psi$ is 
\begin{equation}
V(\bar{\psi},\psi)=\left[\beta_{1}\left(\bar{\psi}\psi\right)^{2}+\beta_{2}\left(i\bar{\psi}\gamma^{5}\psi\right)^{2}\right]^{n},
\label{self_potential}
\end{equation}
where $n$, $\beta_{1}$ and $\beta_{2}$ are constants.  It is interesting to note that this potential introduce  natural  scales in the problem since  the coupling constants $\beta_{1}$ and $\beta_{2}$ have dimension $2(3n-2)$ in units of mass (in four dimension).  It may be of  practical use  in the classical theories and, more important, will have  fundamental implications in a quantum field theory.  

Let us move back to the problem on we are interested and calculate the pressure . Using the derivatives \eqref{derivatives} we see that
\begin{equation}
\frac{1}{2}\bar{\psi}\frac{dV}{d\bar{\psi}}=n\, V(\bar{\psi},\psi)
\end{equation}
and
\begin{equation}
\frac{1}{2}\frac{dV}{d\psi}\psi = s\, n\, V(\bar{\psi},\psi).
\end{equation}
Now, from the equation \eqref{pressure}  we have 
\begin{equation}
p_f=\left[\left(1+s\right)n-1\right]V(\bar{\psi},\psi).
\label{pressure_s}
\end{equation}
Now, it is convenient to separate the analysis in two different cases, namely $s = \pm 1$.

\subsection{Case I - Fermion as a wave function}
If we choose  $s =1$ we  have the case considered  in reference  \cite{Ribas:2005vr} when the fermion is treated as a wave function. From the expression   \eqref{pressure_s} we have the pressure
\begin{equation}
p_f=\left(2n-1\right)V(\bar{\psi},\psi).
\label{pressure_n}
\end{equation}  
According to (\ref{density}), in regimes for which the mass term $m\bar{\psi}\psi$ in \eqref{density} may be neglected when compared  with $V(\bar{\psi},\psi)$ we have $p_f \approx (-1+2n)\rho_f$, {with an equation of state parameter $w_f = -1 + 2n$, which corresponds to a XCDM model when $n\neq 0$}. For $n=0$ we have the usual equation of state  for the vacuum. In general,  the  fermion will have a negative pressure  for   $n < \frac{1}{2}$, {with a phantom behaviour for $n<0$. Such regime must be avoided in order to not violate the weak energy condition (WEC), namely $(\rho + p) >0$}.

The conservation equation (\ref{Eqconserv}) for the massless case is:
\begin{equation}
    \dot{\rho}_f + 6nH\rho_f = 0\,,
\end{equation}
whose solution in terms of the scale factor $a$ is:
\begin{equation}
    \rho_f = \rho_{f,0}\bigg(\frac{a_0}{a}\bigg)^{6n}\,,\label{rhof1}
\end{equation}
and the Friedmann equation that follows from (\ref{EqEinstein}) for a matter energy density plus fermionic field contribution (\ref{rhof1}) is:
\begin{equation}
H^{2}(z)=H_{0}^{2}\bigg[ \Omega_{m} (1+z)^{3}  + \Omega_{f} (1+z)^{6n}  \bigg]. \label{hubble1}
\end{equation}
Here, $H(z)$ and $H_0$ are the Hubble parameter at the redshift $z$ and at present time. $\Omega_{m}\equiv \rho_{m,0}/\rho_{c,0}$ is the present day density parameter for the matter content, $\Omega_{f}\equiv \rho_{f,0}/\rho_{c,0}$ is the density parameter for the fermion field and $\rho_{c,0}=3H_0^2/8\pi G$ is the critical density.

 The last term in \eqref{hubble1} is the contribution of the fermionic field when treated as a wave function.  For $n=0$ we have the ordinary vacuum or cosmological constant density contribution, namely $\Omega_{f} \equiv \Omega_{\Lambda}$.  The free parameters will be constrained with observational data in next section.

\subsection{Case II - Fermion as a  classical field}

{The choice $s=-1$ corresponds to the case where the fermion is a classical field. So, the field $\psi$ is a Grassmann variable. It is very different from the wave function picture used in recent works \cite{Ribas:2005vr,Ribas2007,Rakhi,Vignolo,Inagaki,Channuie,Carloni}, where the result is dependent from the parameter $n$ in the potential $V(\bar{\psi},\psi)$. The independence on the $n$ parameter bring new interesting cosmological consequences, as we will see bellow.}

When $\psi$ is a Grassmann variable the choice $ s = -1$  in the equation  \eqref{pressure_s} gives
\begin{equation}
p_f = - V(\bar{\psi},\psi)\,.\label{Eq26}
\end{equation}
From (\ref{density}) we have the very interesting result that, for a massless fermionic field, the equation of state satisfied by the field is always of vacuum or dark energy type, namely $p_f = - \rho_f$, regardless of the form of the potential. 

For the general case of a massive field the energy density (\ref{density}) can be written in terms of the constant bilinear term associated to the mass, $\rho_f^*=m\bar{\psi}\psi$, namely $\rho_f = \rho_f^* + V(\bar{\psi},\psi)$, and the conservation equation (\ref{Eqconserv}) is:
\begin{equation}
    \dot{\rho}_f + 3H\rho_f^* = 0\,,
\end{equation}
whose solution in terms of the scale factor $a$ is:
\begin{equation}
    \rho_f = 3\rho_f^*\ln \bigg(\frac{a_0}{a}\bigg) + \rho_C\,,\label{rhof}
\end{equation}
and $\rho_C$ is a constant density parameter that must satisfy some boundary condition. For the present time, $a=a_0$, this constant energy density can be associated to the cosmological constant, thus we will take $\rho_C = \rho_\Lambda$. 

The Friedmann equation that follows from (\ref{EqEinstein}) for a standard matter energy density plus fermionic field contribution (\ref{rhof}) is:
\begin{equation}
    H^2 = H_0^2 \bigg[ \Omega_{m}(1+z)^3 + 3\Omega^*\ln(1+z)+\Omega_\Lambda \bigg]\,,\label{H23}
\end{equation}
where $\Omega_m\equiv \rho_{m,0}/\rho_{c,0}$ is the present day matter density parameter, $\Omega^*\equiv \rho_f^*/\rho_{c,0}$ and $\Omega_\Lambda \equiv \rho_\Lambda/\rho_{c,0}$. The last two terms represents an effective time varying dark energy density parameter, {a specific kind of $w(z)$CDM model}, which correctly reproduces the cosmological constant term at $z=0$. Such parameters will be constrained with observational data in next section.

\begin{figure}[h!]
\centering
\includegraphics[width=0.90\linewidth]{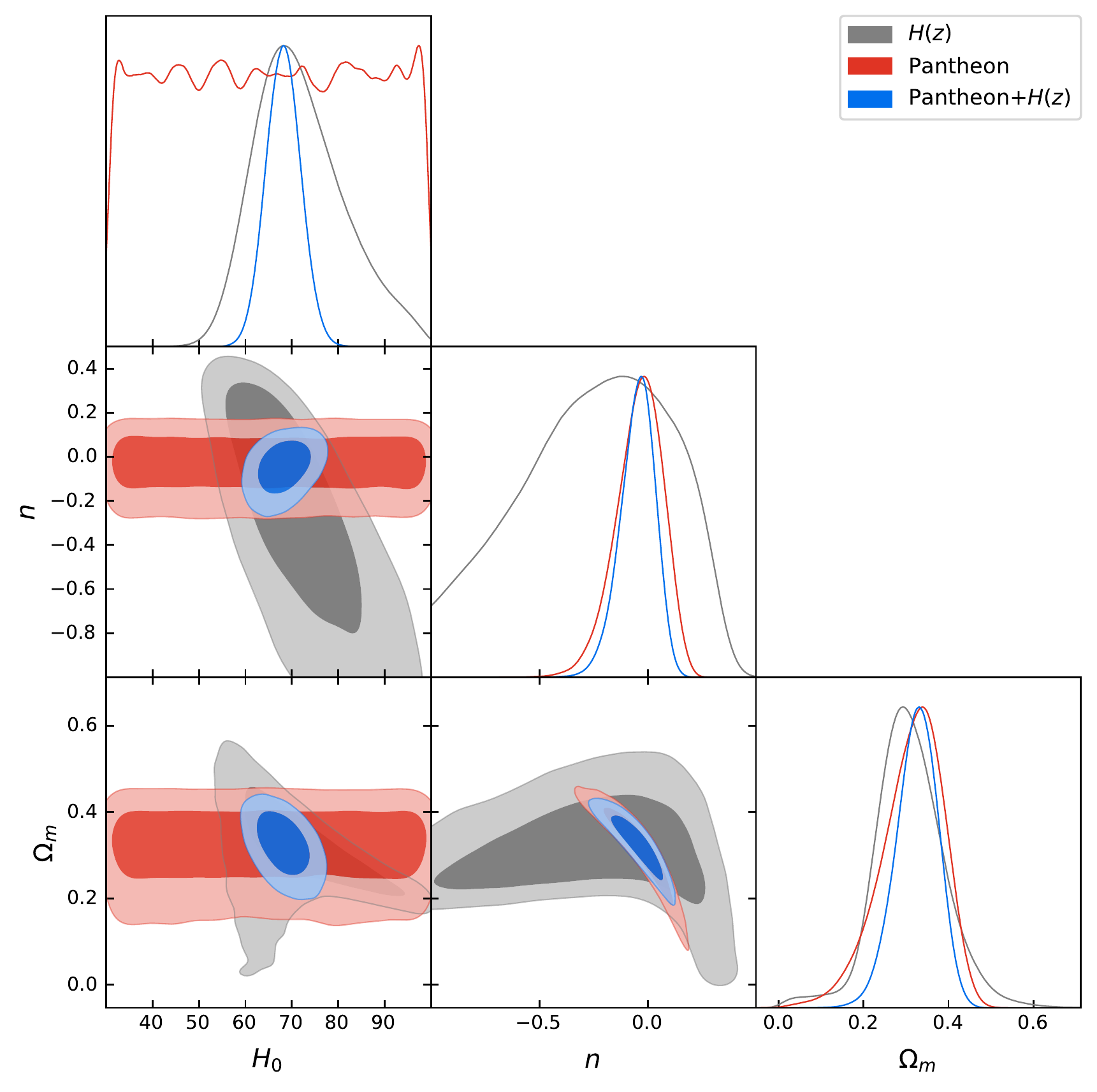}
\caption{Case I - Contours of the parameters $H_0$, $\Omega_m$ and $n$ at $1\sigma$ and $2\sigma$ c.l. for the separate set of data of $H(z)$ (grey) and SNe Ia - Pantheon (red), and for the joint analysis (blue).}
\label{Fig01}
\end{figure}

\begin{figure}[h]
\centering
\includegraphics[width=0.90\linewidth]{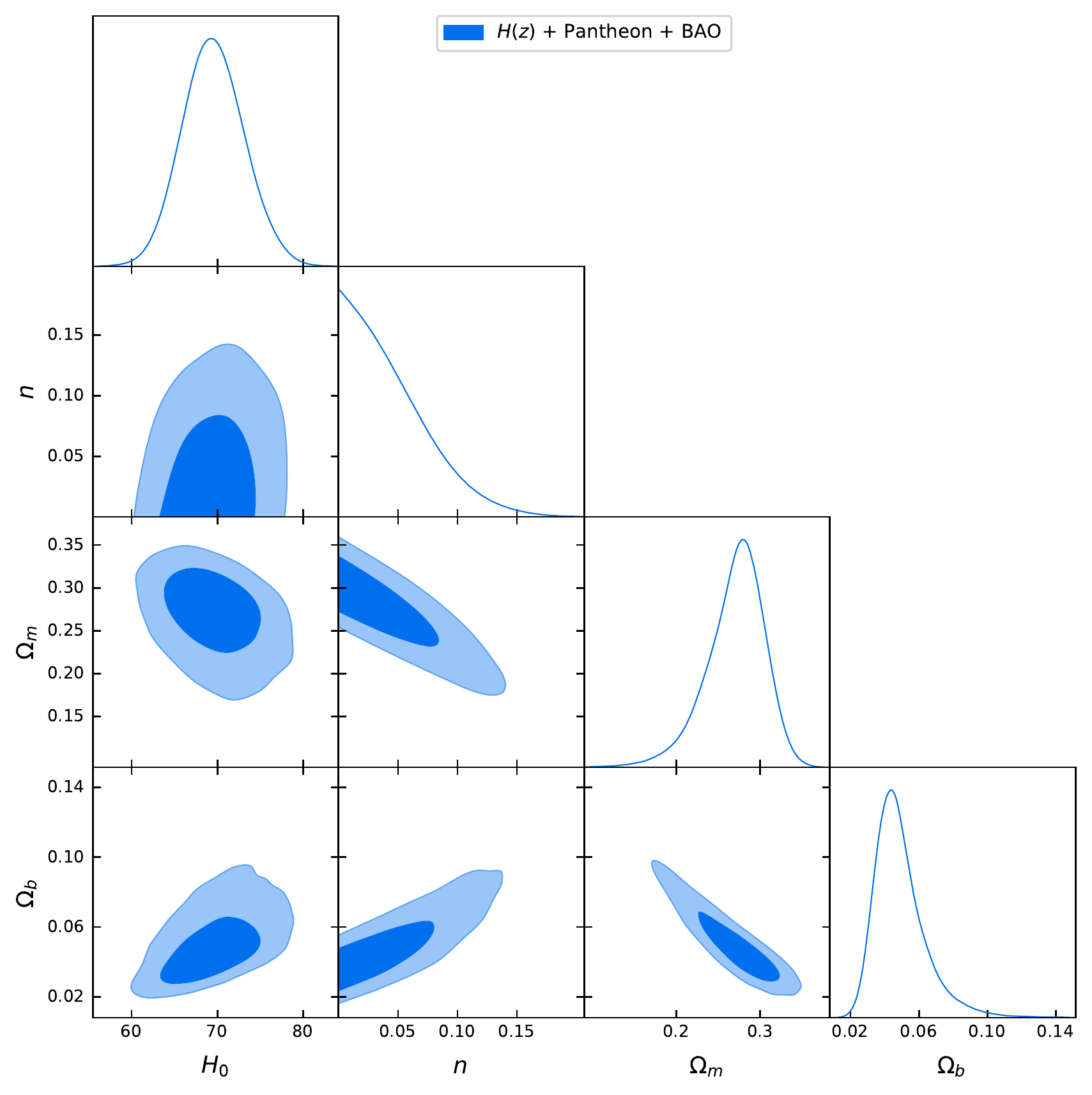}
\caption{Case I - Contours of the parameters $H_0$, $\Omega_m$, $\Omega_b$ and $n$ at $1\sigma$ and $2\sigma$ c.l. for the joint analysis of $H(z)$ + SNe Ia (Pantheon) + BAO, with the prior $n\geq 0$.}
\label{Fig02}
\end{figure}

\begin{table}[h!]
    \centering
    \begin{tabular} { l  c}

 Parameter &  Mean value\\
\hline
\\
{\boldmath$H_0            $} & $69.5^{+3.6 \, +7.7\, +10}_{-3.6\, -7.0\, -10}      $\\

\\

{\boldmath$\Omega_m       $} & $0.270^{+0.039\, +0.066\, +0.086}_{-0.026\, -0.075\, -0.15}   $\\

\\

{\boldmath$\Omega_b       $} & $0.0493^{+0.0083\, +0.033\, +0.089}_{-0.017\, -0.026\, -0.031}   $\\

\\

{\boldmath $n$} &  $0.0^{+0.056 \, +0.115 \, +0.168  }_{-0.0\,-0.0\,-0.0}$\\

\\
\hline
\end{tabular}
    \begin{center}
         \caption{Case I - Mean value of the parameters $H_0 (km\,s^{-1}Mpc^{-1})$, $\Omega_m$, $\Omega_b$  and $n$ from Fig. \ref{Fig02} and 68\%, 95\% and 99\% c.l. limits.}
    \end{center}
 \label{Tab01}
\end{table}

\section{Constraints from observational data}
\label{sec:constraint}

 The analyses to constraint the free parameters of the previous models were implemented in {\sffamily Python} language. The standard method is to
sampling the likelihood $\mathcal{L} \propto e^{-\chi^2/2}$ through Monte Carlo Markov Chain analysis, where 
 the $\chi^2$ function for Pantheon data is given by:
 \begin{equation}
\chi^2_{SN} = \left[\bm{m}_{obs}-{\bm m}(z,\bm{s})\right]^T\bm{C}^{-1}_{SN}\left[\bm{m}_{obs}-{\bm m}(z,\bm{s})\right]\,.
\label{chi2SN}
\end{equation}
 The parameter vector is represented by $\mathbf{s}$, with $\bm{m}_{obs}$ and $\bm{m}$ representing the observed apparent magnitude vector and model apparent magnitude, respectively, and $\bm{C}$ is a covariance matrix containing systematic errors for supernovae data \cite{Betoule}. {We have marginalized the absolute magnitude $M_B$ of SNe Ia over the interval $M_B\in\;[-21,-18]$.} We have used the 1048 Supernovae type Ia data from Pantheon compilation \cite{pantheon}. 
 
 {For the $H(z)$ data we have used the 32 model independent $H(z)$ measurements obtained with Cosmic Chronometers method, obtained from Table I of Moresco {\it et al.} \cite{Moresco2022}. Also, due to the fact that $H(z)$ measurements from galaxy age are strongly biased with several systematic effects \cite{Moresco2022,Kjerr2021, Moresco2020}, we have implemented the covariance matrix effects described in \cite{Moresco2020}, so that the $\chi^2$ function for the $H(z)$ data is:
 \begin{equation}
\chi^2_H = \left[ H_{obs,i} - H(z_i,\mathbf{s})\right]^T\bm{D}^{-1}_{ij}\left[ H_{obs,j} - H(z_j,\mathbf{s})\right] \,,
\end{equation}
where $\bm{D}^{-1}_{ij}$ represents the covariance matrix.}

{For the analysis with BAO data we have used signature estimate
from various sources, as indicated in Tables II and III of Ref. \cite{Camarena2018} and the method described there and in \cite{NunesJesus2020}}.

\subsection{Case I}

In order to study the Case I with the fermionic wave function as a candidate to describe the dark energy in the universe, we start by analysing the model with $H(z)$ and SN Ia - Pantheon data. First, notice that Eq. (\ref{hubble1}) for the present time gives $1=\Omega_m + \Omega_f$, so that $\Omega_f$ can be written as a function of $\Omega_m$ and we are left with a 3 free parameters model, namely $\mathbf{s} = [H_0, \,\Omega_m,\,n]$. {The parameter $n$ characterizes the deviation from the standard $\Lambda$CDM model.}

{The contours at $1\sigma$ and $2\sigma$ c.l. for the parameters $H_0$, $\Omega_m$ and $n$ for the separate set of data are shown in Fig. \ref{Fig01}, with $H(z)$ (grey), SNe Ia - Pantheon (red) and the combined analysis (blue). The mean values of the parameters for the joint analysis of $H(z)$ + SN Ia is $n=-0.05^{+0.16}_{-0.17}$, $\Omega_{m} = 0.325^{+0.096}_{-0.10}$ and $H_0 = 68.2^{+7.5}_{-7.3}$ at 95\% c.l..  In this first analysis no prior over $n$ was imposed, and we can see that a negative value is favoured, which indicates a violation of the weak energy condition. Also, it is well known that $H(z)$ and Supernovae data does not constraint the value of the barion density parameter, $\Omega_b$. Thus, in a second analysis we add the BAO data set and put a prior on the $n$ parameter, namely $n\geq 0$. The contours at $1\sigma$ and $2\sigma$ c.l. are displayed in Fig. \ref{Fig02} for the joint analysis of $H(z)$ + SNe Ia (Pantheon) + BAO. The mean values and 95\% c.l.  limits for each parameter are displayed in Table I. We can see that the values of the parameters $H_0$ and $\Omega_m + \Omega_b$ are in full agreement to the values obtained for the $\Lambda$CDM based model from Planck 2018 latest results \cite{Planck2018} ($H_0=(67.4\pm 0.5)$km/s/Mpc and  $\Omega_{m}=0.315\pm{0.007}$ at 68\% c.l.). The small positive value of the $n$ parameter indicates the deviation from the standard $\Lambda$CDM model, which corresponds to $n=0$. In next Section we study the linear evolution of density perturbations for this model.
}

\subsection{Case II}
 
For this case, Eq. (\ref{H23}) gives $1=\Omega_m + \Omega_\Lambda$ at present time, so that $\Omega_\Lambda$ can be written as a function of $\Omega_m$ and we are left with a 3 free parameters model, namely $\mathbf{s} = [H_0, \,\Omega_m,\,\Omega^*]$. {The parameter $\Omega^*$ characterizes the deviation from the standard $\Lambda$CDM model.}

\begin{figure}[t]
\centering
\includegraphics[width=0.90\linewidth]{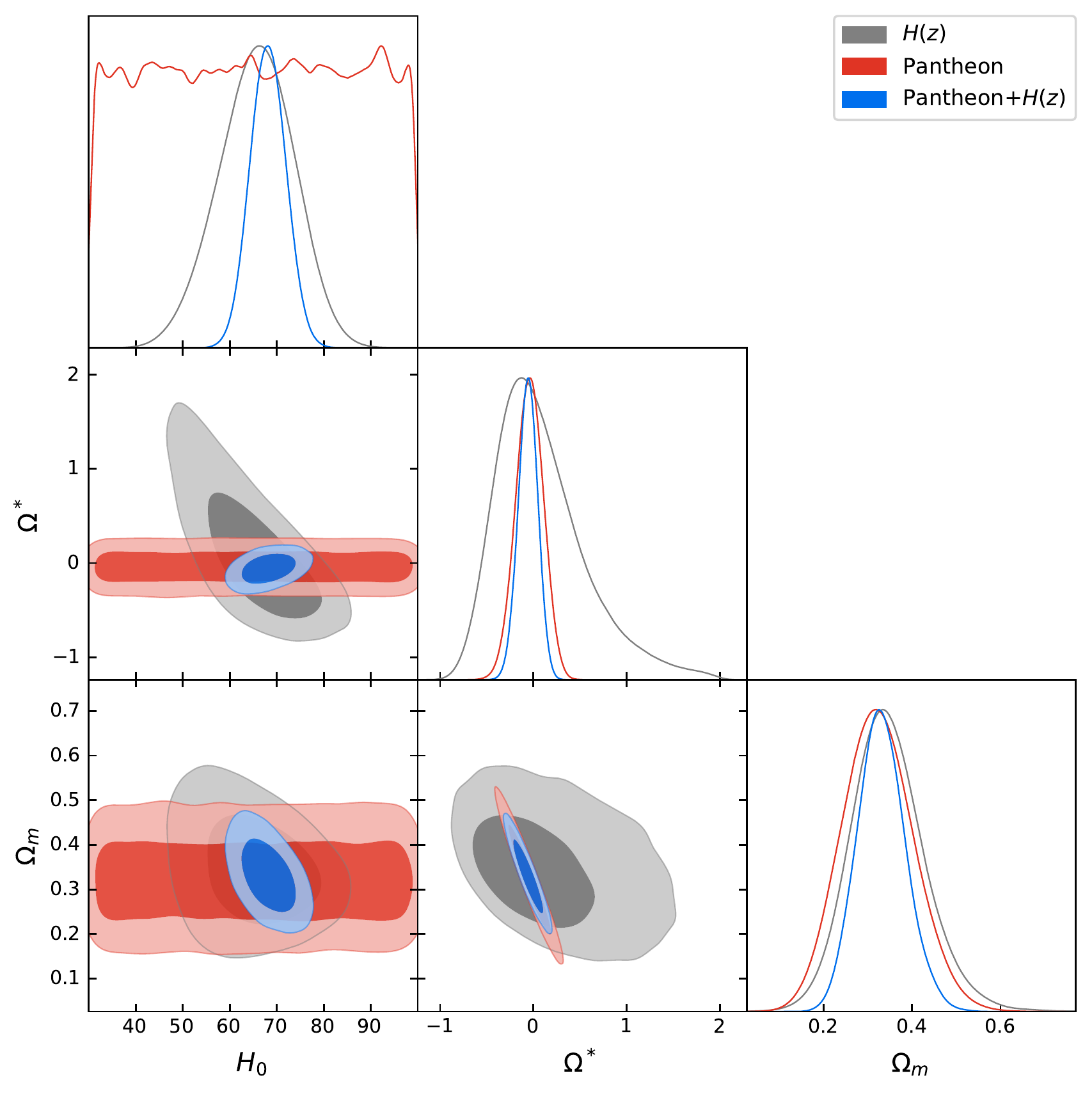}
\caption{Case II - Contours of the parameters $H_0$, $\Omega_m$ and $\Omega^*$ at $1\sigma$ and $2\sigma$ c.l. for the separate set of data of $H(z)$ (grey), SNe Ia - Pantheon (red) and the joint analysis (blue).}
\label{Fig03}
\end{figure}

\begin{figure}[h]
\centering
\includegraphics[width=0.90\linewidth]{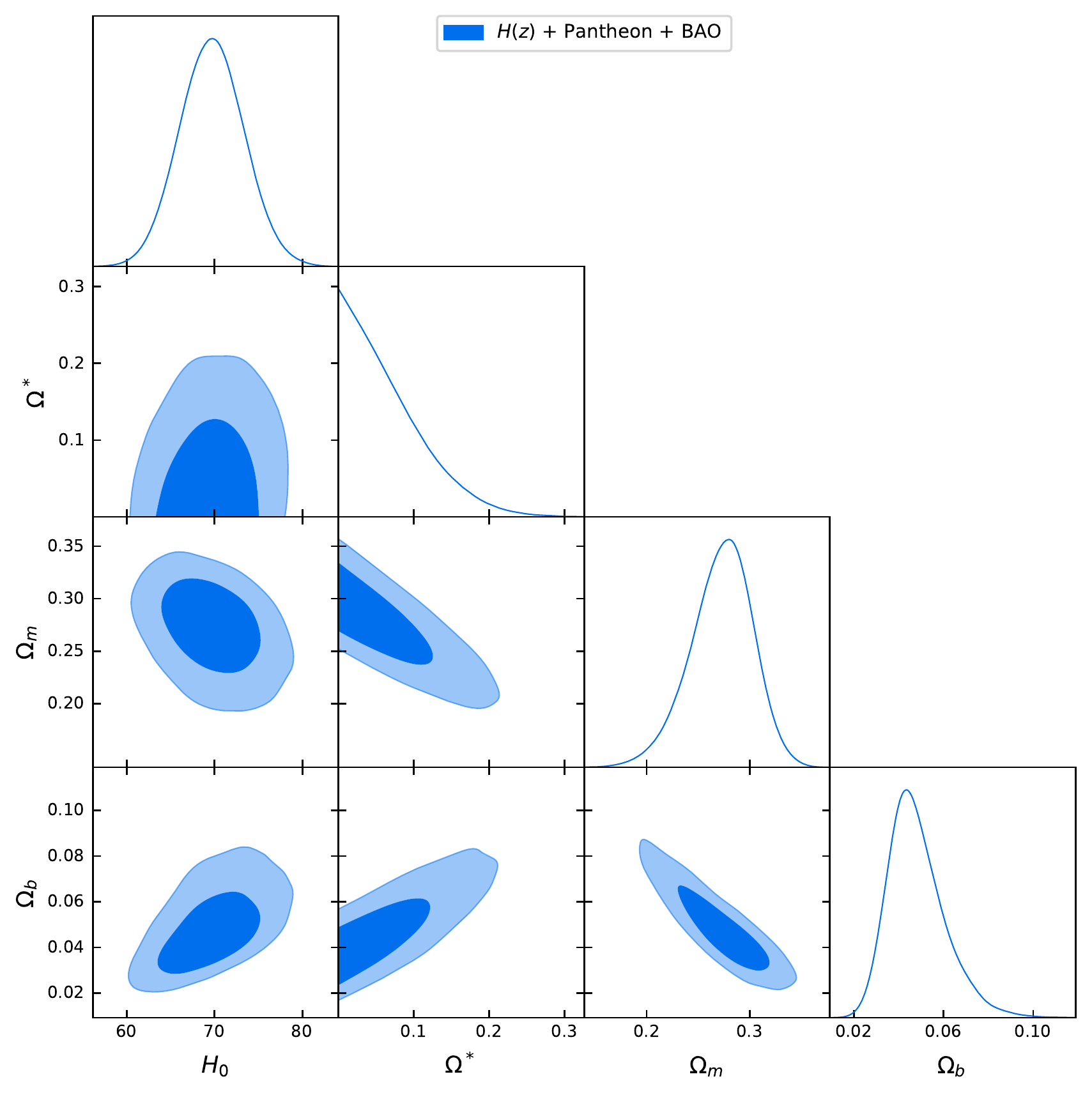}
\caption{Case II - Contours of the parameters $H_0$, $\Omega_m$, $\Omega_b$ and $\Omega^*$ at $1\sigma$ and $2\sigma$ c.l. for the joint analysis of $H(z)$ + SNe Ia (Pantheon) + BAO, with the prior $\Omega^* \geq 0$.}
\label{Fig04}
\end{figure}

{We start the analysis with $H(z)$ and SNe Ia - Pantheon data. The contours at $1\sigma$ and $2\sigma$ c.l. for the parameters $H_0$, $\Omega_m$ and $\Omega^*$ are shown in Fig. \ref{Fig03}, with $H(z)$ (grey), SNe Ia (red) and the joint analysis of $H(z)$ + SNe Ia (blue). The mean values of the parameter are $\Omega^*=-0.06^{+0.21}_{-0.21}$, $\Omega_{m} = 0.33^{+0.12}_{-0.10}$ and $H_0 = 68^{+8}_{-7}$ at 95\% c.l..  As in the first case, no prior over $\Omega^*$ was imposed initially, and we can see that a negative value is allowed for the observational data. However such parameter is directly related to the mass of the fermionic field through $\Omega^*_f=\rho^*_f/\rho_{c,0}$, with $\rho^*_f = m \bar{\psi}\psi$. Thus, in a second analysis we impose the prior $\Omega^* \geq 0$ and add the BAO data set. The contours at $1\sigma$ and $2\sigma$ c.l. are displayed in Fig. \ref{Fig04} for the joint analysis of $H(z)$ + SNe Ia (Pantheon) + BAO. The mean values and 95\% c.l.  limits for each parameter are displayed in Table II. The values of the parameters $H_0$ and $\Omega_m + \Omega_b$ are in full agreement to the values obtained for the $\Lambda$CDM based model from Planck 2018 results \cite{Planck2018}. The small positive value of the $\Omega^*$ parameter indicates the deviation from the standard $\Lambda$CDM model, which corresponds to $\Omega^*=0$. In next Section we study the linear evolution of density perturbations for this model.
}

\begin{table}[h!]
    \centering
    \begin{tabular} { l  c}

 Parameter &  Mean value\\
\hline
\\
{\boldmath$H_0            $} & $69.7^{+3.7 \, +7.4\, +10.0}_{-3.7\, -7.2\, -10.0}      $\\

\\

{\boldmath$\Omega_m       $} & $0.272^{+0.033\, +0.056\, +0.080}_{-0.026\, -0.068\, -0.11}   $\\

\\

{\boldmath$\Omega_b       $} & $0.048^{+0.009\, +0.027\, +0.057}_{-0.015\, -0.023\, -0.029}   $\\

\\

{\boldmath$\Omega^*       $} & $0.0^{+0.084\, +0.176\, +0.262}_{-0.0\, -0.0\, -0.0}   $\\

\\
\hline
\end{tabular}
    \begin{center}
         \caption{Case II - Mean value of the parameters $H_0 (km\,s^{-1}Mpc^{-1})$, $\Omega_m$ , $\Omega_b$ and $\Omega^*$ from Fig. \ref{Fig04} and 68\%, 95\% and 99\% c.l. limits.}
    \end{center}
 \label{Tab02}
\end{table}

\section{Linear evolution of density perturbations}

As a final analysis, let us study the linear evolution of density perturbations related to the models of Cases I and {II, where a non null value of the parameters $n$ and $\Omega^*$ represents the deviation from $\Lambda$CDM model. For both cases we have an effective equation of state parameter $w_f$ whose} effects on the primordial density perturbations can be studied using the techniques developed by Abramo {\it et al.} \cite{abramo}, where structure formation in the presence of dark energy perturbations are considered in the so called pseudo-Newtonian approach.

Written the conservation equation (\ref{Eqconserv}) as:
\begin{equation}
\frac{d\rho_f}{dz} = \frac{3(1+w_f)}{(1+z)}\rho_f\,,\label{rhoz}
\end{equation}
{and using (\ref{rhof1}) and (\ref{rhof}) for Cases I and II, respectively, we obtain:
\begin{equation}
    w_f = -1 + 2n\,, \hspace{1cm} \text{Case I}\label{wf1}
\end{equation}
\begin{equation}
    w_f = -1 + \frac{\Omega^*}{\Omega_f(z)}\,,\hspace{1cm} \text{Case II}\label{wf}
\end{equation}
where $\Omega_f(z)=3\Omega^* \ln (1+z)+\Omega_\Lambda$ represents an effective dark energy density parameter which evolves with time, according to (\ref{H23}). At this point it becomes clear that Case I represents a XCDM model with a constant equation of state parameter and Case II represents a $w(z)$CDM model, with a varying equation of state parameter.}

\begin{figure}[t!]
\centering
\includegraphics[width=0.80\linewidth]{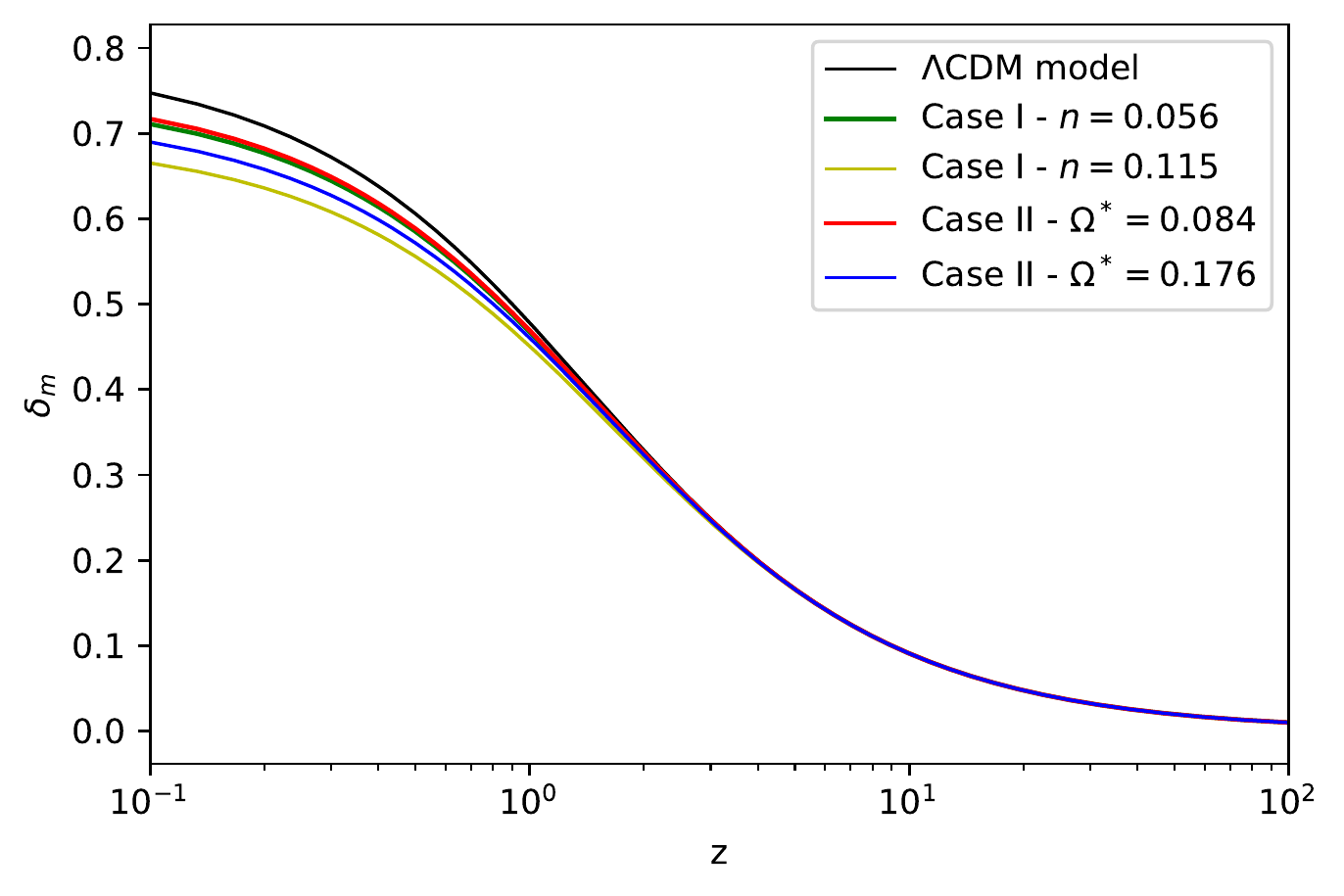}\\
\includegraphics[width=0.80\linewidth]{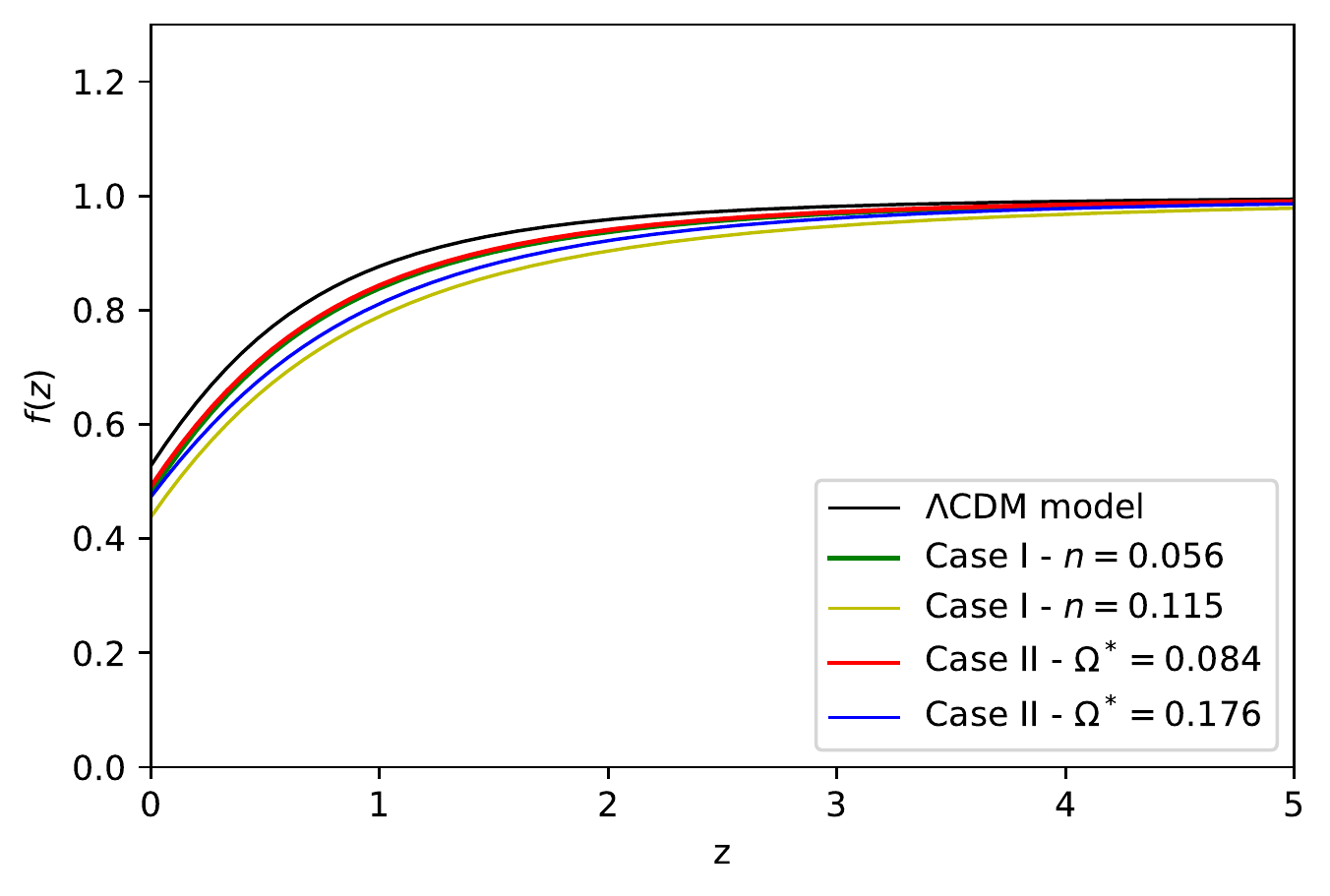}
\caption{{\bf Top panel} - Growth of the density contrast $\delta_m$ of matter for both cases. The curves were plotted with the values of the parameters in Tables I and II, with different values of $n$ and $\Omega^*$. The evolution of matter density contrast for $\Lambda$CDM model is also presented for comparison (upper black line). {\bf Bottom panel} - Growth function $f$ of matter for the fermionic model and for $\Lambda$CDM model (black line) for comparison.}
\label{Fig05}
\end{figure}

The cosmological perturbations can be introduced by admitting inhomogeneous deviations from the background quantities,
\begin{equation}
\rho_m=\bar{\rho}_{m}(1+\delta_m),\quad p_m=\bar{p}_m+\delta p_m,\quad\rho_f=\bar{\rho}_{f}(1+\delta_f),\quad p_f=\bar{p}_f+\delta p_f,
\end{equation}
and therefore $\delta_m=\delta\rho_m(\vec{x},t)/\bar{\rho}_{m}$ and $\delta_f=\delta\rho_f(\vec{x},t)/\bar{\rho}_{f}$ are the density contrasts for each fluid.
 The linear regime is described by the coupled differential equations (see Eqs. (29) and (30) of \cite{abramo}):
\begin{equation}
\ddot{\delta}_m+2H\dot{\delta}_m=\frac{3H^2}{2}[\Omega_m(t)\delta_m + \Omega_f(t) \delta_f (1+3w_f)]\,,\label{deltam}
\end{equation}
\begin{equation}
\ddot{\delta}_f+\bigg(2H-\frac{\dot{w}_f}{1+w_f}\bigg)\dot{\delta}_f=\frac{3H^2}{2}(1+w_f)[\Omega_m(t)\delta_m + \Omega_f(t) \delta_f (1+3w_f)]\,,\label{deltaphi}
\end{equation}
where a dot stands for time derivative and the density parameters $\Omega_m(t)=\Omega_m(a(t))$ and $\Omega_f(t)=\Omega_f(a(t))$ are functions of time at this level, but can be rewritten as a function of the redshift in the standard manner, using $\frac{d}{dt}=-(1+z)H(z)\frac{d}{dz}$.

{The solution of the system of coupled differential equations (\ref{deltam})-(\ref{deltaphi}) can be obtained numerically with the initial conditions at $z_i = 10^3$, namely $\delta_{m}(z_i)=1/(1+z_i)$ and $\delta'_{m}(z_i)=-\delta_{m}(z_i)/(1+z_i)$, where a prime denotes derivative with respect to the redshift. The initial conditions of the dark energy density are assumed to be null.
}

{In Figure \ref{Fig05} we show the effects of fermionic field fluctuations on the growth of dark
matter perturbation $\delta_m$ for both cases with different values of $n$ and $\Omega^*$. The $\Lambda$CDM model is also present for comparison. On the top panel we see that the effect of the dark energy perturbation represented by the fermionic field is to diminish the growth of the matter perturbation for late time. For the best fit values of $n$ and $\Omega^*$, green and red curves respectively, the growth of matter perturbation are quite indistinguishable each other. For $2\sigma$ the models are slightly different, yellow and blue curves, respectively. }

{The growth function $f(z)$ defined by $f=-(1+z)\frac{\delta'_m}{\delta_m}$ is showed in the bottom panel. The main behaviours are quit similar to the growth of density contrast. Although being indistinguishable from $\Lambda$CDM model (black line) at early time, the fermionic model predicts the formation of fewer structures over the evolution for late time, as expected when compared to standard model. Such qualitative behaviour could alleviate at least one of the so called "small scale problems" \cite{Nakama2017} associated to $\Lambda$CDM model, which predict too many dwarf galaxies and too much dark matter in the innermost regions of galaxies. The presence of an overdensity of dark energy in the same regions of dark matter density perturbations could diminish the net effect of dark matter.}

\section{Conclusion}\label{sec 6}

In this paper we have investigated the late time cosmological evolution driven by a fermionic field endowed with a potential written in terms of
scalar and pseudoscalar invariants. Two slightly different scenarios are possible, namely, whether the fermion field is considered a standard wave-function or a grassmanian variable. In both approaches the energy density and pressure of the field can be treated as effective thermodynamic quantities that enters the Friedmann evolution equation satisfying an equation of state of the form $p_f \sim w \rho_f$, with $w$ an effective equation of state parameter.

The free parameters of the model composed by a total matter energy density $\rho_m$ and fermion field energy density $\rho_f$ were constrained with Supernovae and Hubble parameter observational data. {The inclusion of BAO data into the analysis allows the constraint of barionic density parameter separate out of the total matter density.} The $H_0$ parameter was also set as a free parameter in order to compare with the recent Planck 2018 results.

In the first approach, where a massless fermionic field is treated as a wave-function, the equation of state can be written as $p_f = (-1+2n)\rho_f$. {Such model is equivalent to a XCDM model with constant equation of state parameter.} The free parameter $n$, which represents the deviation of the model from the $\Lambda$CDM model, was constrained by observational data, furnishing $n<0.115$ at $2\sigma$ c.l.. The values of $H_0$, $\Omega_m$ and $\Omega_m$ (see Table I) are also in good agreement to latest Planck 2018 results \cite{Planck2018}.

In the second approach, where a massive fermionic field is treated as a grassmanian variable, the equation of state can be written as $p_f = -\rho_f + \rho_f^*$, where the term $\rho_f^* = m\bar{\psi}\psi$ represents a deviation from a vacuum equation of state parameter, depending on the mass $m$ of the fermionic field. {The model is equivalent to a $w(z)$CDM model.} For $m=0$ the system represents the $\Lambda$CDM model. The constraint with observational data gives the value $\Omega^* < 0.176$ at $2\sigma$ c.l..  The values of the parameters $H_0$, $\Omega_m$ and $\Omega_b$ (see Table II) are also in good agreement to latest Planck 2018 results \cite{Planck2018}.

Both analyzes show that the fermionic field with a potential written in terms of bilinear invariants of the form \eqref{self_potential} are in good agreement with the latest Planck 2018 results for the $\Lambda$CDM model, representing an extension of the standard model and given a new interpretation for the dark energy component. For the specific case of the second approach, where the fermionic field is a Grassmann variable, a much more interesting feature appears. If we neglect the mass $m$ of the fermion in \eqref{density}, Eq. (\ref{Eq26}) ensures that the fermion behaves like the vacuum energy for any potential of the form \eqref{self_potential}. This shows that the intrinsic grassmannian property of the field leads always to a dark energy or negative pressure contribution on the evolution, irrespective the specific form of the potential. From a quantum point of view this feature may be related to the degeneracy pressure intrinsic to the genuine fermion field.

Finally, the study of linear evolution of density perturbations for both cases show that, at least qualitatively, the problem of overdensities predicted by $\Lambda$CDM model during the formation of the first structures can be alleviated for non negative values of the parameters $n$ and $\Omega^*$, since that for both cases the growth function $f$ are smaller than for the $\Lambda$CDM model. {Additionally, as extensions of the standard model, further investigations on the $H_0$ tension and $\sigma_8$ problem \cite{Efstathiou2021,Camarena2021,Nunes2021,Alam2021} must be carried out for both cases}.

\begin{acknowledgements}
This study was financed in part by the Coordena\c{c}\~ao de Aperfei\c{c}oamento de Pessoal de N\'ivel Superior - Brasil (CAPES) - Finance Code 001. SHP would like to thank CNPq - Conselho Nacional de Desenvolvimento Cient\'ifico e Tecnol\'ogico, Brazilian research agency, for financial support, grants numbers 303583/2018-5 and 308469/2021-6. 
\end{acknowledgements}


\begin{thebibliography}{99}

\bibitem{peebles}
P. J. E. Peebles, "Cosmology’s century: an inside history of our modern understanding of the universe". Princeton University Press, 2020.

\bibitem{oks}
E. Oks,
New Astronomy Reviews 93, 101632 (2021).

\bibitem{Arun}
K. Arun, S. B. Gudennavar, and C. Sivaram, 
Advances in Space Research 60, 166-186 (2017).

\bibitem{Ribas:2005vr}
M.~O.~Ribas, F.~P.~Devecchi and G.~M.~Kremer,
Phys. Rev. D \textbf{72}, 123502 (2005) [arXiv:gr-qc/0511099 [gr-qc]].

\bibitem{Ribas2007}M. O. Ribas, F. P. Devecchi, G. M. Kremer, Europhys. Lett. 81, 19001, (2008); [arXiv:0710.5155 [gr-qc]]

\bibitem{Magana2018}
J.~Magana, M.~H.~Amante, M.~A.~Garcia-Aspeitia and V.~Motta,
Mon. Not. Roy. Astron. Soc. \textbf{476} (2018) no.1, 1036-1049,
[arXiv:1706.09848 [astro-ph.CO]].

\bibitem{Moresco2022} M. Moresco et. al.,
[arXiv:2201.07241 [astro-ph.CO]].

\bibitem{pantheon}
  D.~M.~Scolnic {\it et al.},
  Astrophys.\ J.\  {\bf 859} (2018) no.2,  101
  [arXiv:1710.00845 [astro-ph.CO]].

\bibitem{Planck2018} Planck Collaboration: N. Aghanim et al. 
Astron. Astrophys.  641, A6 (2020), [arXiv:1807.06209].

\bibitem{nambu}
Y. Nambu and G. Jona-Lasinio,
Physical review 122,  345 (1961); 
124, 246 (1961).

\bibitem{klevansky}
S. P. Klevansky, 
Reviews of Modern Physics 64, 649 (1992).

\bibitem{donoghue}
J. F. Donoghue, E. Golowich and B. R. Holstein, "Dynamics of the Standard Model", Cambridge University Press, 1996.

\bibitem{Joyce:2014kja}
A.~Joyce, B.~Jain, J.~Khoury and M.~Trodden,
Phys. Rept. \textbf{568} (2015), 1-98.

\bibitem{Steinwachs}
C. F. Steinwachs "Higgs field in cosmology.", One Hundred Years of Gauge Theory, Springer, 253-287 (2020);	arXiv:1909.10528 [hep-ph].

\bibitem{Maleknejad}
A. Maleknejad, M. M. Sheikh-Jabbari and J. Soda, 
Physics Reports 528, 161-261 (2013).

\bibitem{kolbe}
E. Kolbe and M. Turner, "The early universe", Frontiers in Physics, Westview Press, 1994.

\bibitem{Susskind}
L. Susskind, 
Physics Reports 104, 181-193 (1984).

\bibitem{Ferretti:2013kya}
G.~Ferretti and D.~Karateev,
JHEP \textbf{03} (2014), 077

\bibitem{Barnard:2013zea}
J.~Barnard, T.~Gherghetta and T.~S.~Ray,
JHEP \textbf{02} (2014), 002.

\bibitem{Csaki:2015hcd}
C.~Csaki, C.~Grojean and J.~Terning,
Rev. Mod. Phys. \textbf{88} (2016) no.4, 045001.

\bibitem{Rakhi}R. Rakhi, G.V. Vijayagovindan, K. Indulekha,
Int. J. Mod. Phys. A25, 2735 (2010); [arXiv:0912.1222v1 [gr-qc]].

\bibitem{Vignolo}S. Vignolo, S. Carloni, L. Fabbri,
Phys. Rev. D 91, 043528 (2015); [arXiv:1412.4674v2 [gr-qc]].

\bibitem{Inagaki}T. Inagaki, S. D. Odintsov, H. Sakamoto, 
Astrophys Space Sci 360, 67 (2015); [	arXiv:1509.03738 [hep-th]].

\bibitem{Channuie}P. Channuie, C. Xiong,
Phys. Rev. D 95, 043521 (2017); [arXiv:1609.04698v2 [hep-ph]].


\bibitem{Carloni}S. Carloni, R. Cianci, P. Feola, E. Piedipalumbo, S.Vignolo,
JCAP 09, 014 (2019); [arXiv:1811.10300 [astro-ph.CO]].




\bibitem{Kremer2003}G. M. Kremer and F. P. Devecchi, Phys. Rev. D 66, 063503
(2002); 67, 047301 (2003); G. M. Kremer, Phys. Rev. D
68, 123507 (2003); Gen. Relativ. Gravit. 35, 1459 (2003).

\bibitem{Betoule}M. Betoule et.al., A\&A 568, A22 (2014), 	[arXiv:1401.4064 [astro-ph.CO]].

\bibitem{Kjerr2021}
A. A. Kjerrgren, E. Mortsell, [arXiv:2106.11317 [astro-ph.CO]].

\bibitem{Moresco2020}M. Moresco et. al., ApJ 898, 82, (2020), [arXiv:2003.07362 [astro-ph.GA]].

\bibitem{Camarena2018}D. Camarena, V. Marra, Phys. Rev. D 98(2), 023537 (2018),
[arXiv:1805.09900 [astro-ph.CO]].

\bibitem{NunesJesus2020}R. C. Nunes, S. K. Yadav, J. F. Jesus, A. Bernui, 
MNRAS, 497, (2020), 2133, [arXiv:2002.09293].

\bibitem{abramo} L. R. Abramo, R. C. Batista, L. Liberato, R. Rosenfeld, 
\textit{JCAP} 0711 (2007) 012, [arXiv:0707.2882 [astro-ph]].



\bibitem{Nakama2017}T. Nakama, J. Chluba, M. Kamionkowski
Phys. Rev. D 95 (12) 121302 (2017), [arXiv:1703.10559]. 

\bibitem{Efstathiou2021} G. Efstathiou, 
MNRAS 505, (2021), 3866, [arXiv:2103.08723].

\bibitem{Camarena2021}D. Camarena, V. Marra, 
MNRAS, 504, (2021), 5164, [arXiv:2101.08641].

\bibitem{Nunes2021}R. C. Nunes and E. Di Valentino,
Phys. Rev. D 104, (2021) 063529, [arXiv:2107.09151].

\bibitem{Alam2021}Shadab Alam et al., Phys. Rev. D 103, 083533 (2021), [arXiv:2007.08991].

\end{thebibliography}


\end{document}